\documentclass[12pt]{iopart}
\usepackage{setstack,latexsym,float,color}
\usepackage{graphicx}

\usepackage{iopams}  
\begin{document}

\title[Luria-Delbr{\"u}ck, revisited]{Luria-Delbr{\"u}ck, revisited:
  The classic experiment does not rule out Lamarckian evolution}

\author{Caroline Holmes$^{1, 2}$, Mahan Ghafari$^1$, Abbas Anzar$^3$, Varun Saravanan$^3$, Ilya Nemenman$^{1, 2}$}
\address{$^1$ Department of Physics, Emory University, Atlanta, GA 30322, USA}
\address{$^2$ Department of Biology, Emory University, Atlanta, GA 30322, USA}
\address{$^3$ Neuroscience  Program, Emory University, Atlanta, GA 30322, USA}
\ead{ilya.nemenman@emory.edu}
\vspace{10pt}

\begin{abstract}
  We re-examined data from the classic Luria-Delbr\"uck fluctuation
  experiment, which is often credited with establishing a Darwinian
  basis for evolution. We argue that, for the Lamarckian model of
  evolution to be ruled out by the experiment, the experiment must
  favor pure Darwinian evolution over both the Lamarckian model and a
  model that allows both Darwinian and Lamarckian mechanisms. Analysis of the combined model was not performed in the original 1943
  paper. The Luria-Delbr\"uck paper also did not consider the
  possibility of neither model fitting the experiment. Using Bayesian
  model selection, we find that the Luria-Delbr\"uck experiment, indeed,
  favors the Darwinian evolution over purely Lamarckian. However, our
  analysis does not rule out the combined model, and hence cannot rule
  out Lamarckian contributions to the evolutionary
  dynamics.\end{abstract}

%\noindent{\it Keywords}: Darwinian evolution, Lamarckian evolution,
%and Bayesian model selection

%\noindent{\it Keywords}: XXXXXX, YYYYY, ZZZZZZZZZ
\submitto{\PB}
%\maketitle

\maketitle

\section{Introduction}

From the dawn of evolutionary biology, two general mechanisms,
Darwinian and Lamarckian, have been routinely considered as
alternative models of evolutionary processes.  The {\em Darwinian
  hypothesis} posits that adaptive traits arise continuously over time
through spontaneous mutation, and that evolution proceeds through
natural selection on this already existing variation. In contrast, the
{\em Lamarckian hypothesis} proposes that adaptive mutations arise in
response to environmental pressures. The Nobel Prize winning
fluctuation test by Salvador Luria and Max Delbr\"uck
\cite{luria1943mutations} is credited with settling this debate, at
least in the context of evolution of phage-resistant bacterial cells.

\begin{figure}[!h] \includegraphics[width=14cm]{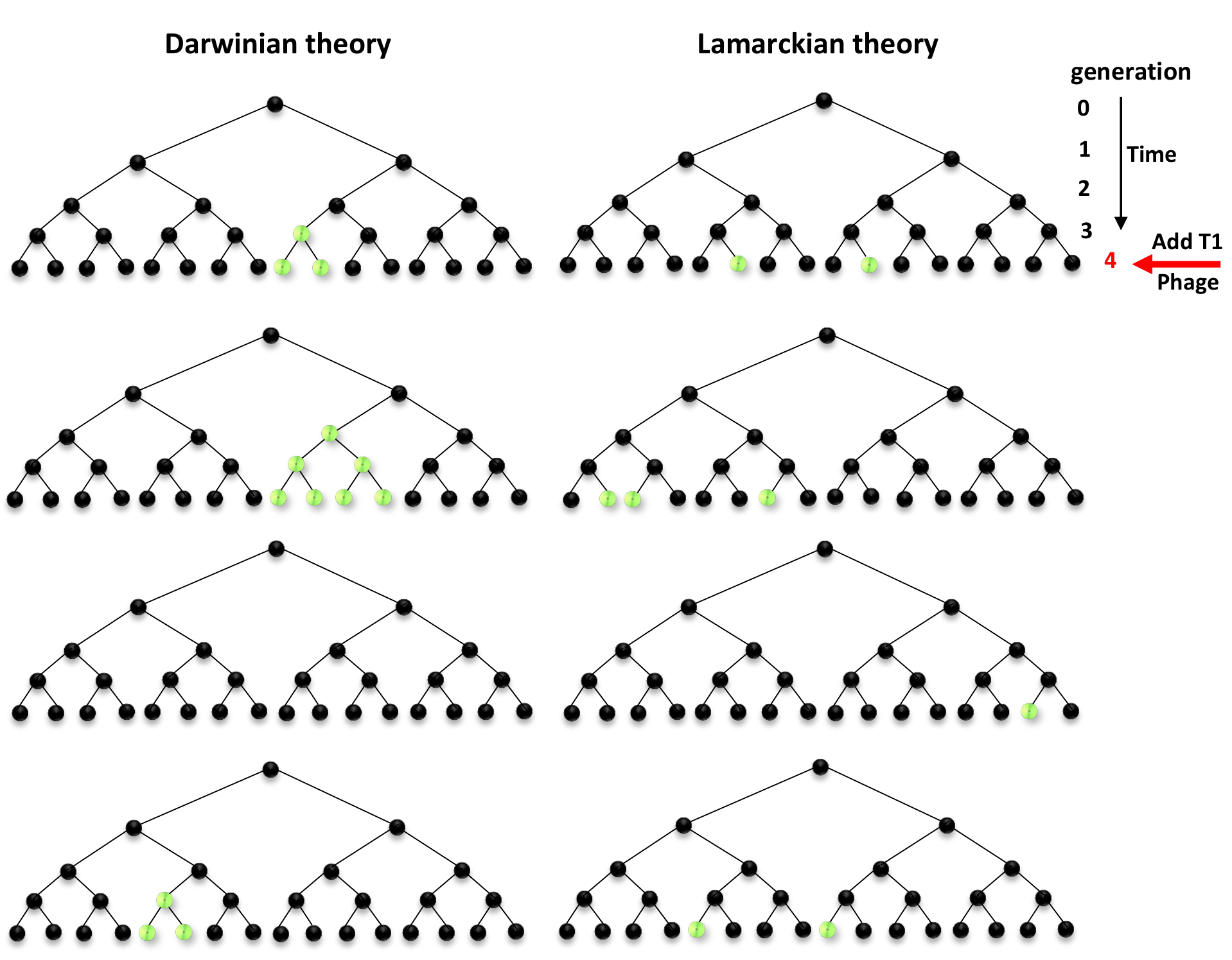} \centering
  \caption{Schematics of the two theories tested by the
    Luria-Delbr\"uck experiment. Here black dots denote bacteria
    susceptible to the bacteriophage, and green dots resistant
    bacteria. Each tree represents one realization of the experiment,
    which starts with a single bacterium (top). The bacterium then
    divides for several generations, and the phage is introduced into
    the culture at the last generation (bottom row of each
    tree). Darwinian theory (left column) of evolution predicts that
    mutations happen spontaneously throughout the experiment. Thus
    different repeats of the experiment (different trees) will produce
    very broadly distributed numbers of survivors (from 0 to 4 in this
    example). In contrast, in the Lamarckian case (right column),
    mutations only occur when the phage is introduced, so that the
    standard deviation of the number of survivors in different repeats
    (from 1 to 3 in this example) scales as the square root of their
    mean, which is much smaller than in the Darwinian case.}
\label{fig:tree}
\end{figure}

Luria and Delbr\"uck realized that the two hypotheses would lead to
different variances (even with the same means) of the number of
bacteria with any single adaptive mutation. Specific to the case of
bacteria exposed to a bacteriophage, this would result in different
distributions of the number of surviving bacteria,
cf.~Fig.~\ref{fig:tree}. In the Darwinian scenario, there is a
possibility of a phage-resistance mutation arising in generations
prior to that subjected to the phage. If this mutation happens many
generations earlier, there will be a large number of resistant progeny
who will survive (a ``jackpot'' event). However, there will be no
survivors if the mutation does not exist in the population at the
moment the phage is introduced. If the same experiment were repeated
many times, the variance of the number of survivors would be large. In
contrast, in the Lamarckian scenario, the distribution of the number
of survivors is Poisson. Indeed, each occurring mutation (and hence each
survivor) happens with a small probability, independent of the
others. This would result in the usual square-root scaling of the
standard deviation of the number of survivors, a much smaller spread
than in the Darwinian case.

To test this experimentally, Luria and Delbr\"uck let the cells grow
for a few generations, exposed them to a phage, plated the culture,
and then counted the number of emergent colonies, each started by a
single resistant, surviving bacterium. They found that the
distribution of the number of survivors, as measured by the number of
colonies grown after plating, was too heavy-tailed to be consistent
with the Poisson distribution. They concluded then that the bacteria
must evolve using the Darwinian mechanism. They could not derive an
analytical form of the distribution of survivors in the Darwinian
model, so that their data analysis was semi-quantitative at best. In
particular, they could only establish that the Darwinian model fits
the data better than the Lamarckian/Poissonian one, but they could not
quantify {\em how good} the fit is. 

Potentially even more importantly, the original paper contrasted only
two scenarios: pure Lamarckian and pure Darwinian ones.  However, it
is possible that both processes have a role in bacterial evolution, as
is abundantly clear now in the epoch of CRISPR and epigenetics
\cite{koonin2009evolution,
  jablonka2002changing,jablonka2009transgenerational, Barrangou2007crispr}.  Ruling out a significant
Lamarckian contribution to evolution would require us to show not only
that the Darwinian model explains the data better than the Lamarckian
one, but also that the Darwininan model is more likely than the {\em
  Combination model}, which allows for both types of evolutionary
processes.  Evolution could also proceed
through an entirely different mechanism, so that neither of the
proposed models explain the data. Distinguishing between these possibilities requires
evaluating whether a specific model fits the data well, rather than
which of the models fits the data better.

Unlike Luria and Delbr\"uck in 1943, we have powerful computers and
new statistical methods at our disposal. Distributions that cannot be
derived analytically can be estimated numerically, and model
comparisons can be done for models with different numbers of
parameters. In this paper, we perform the quantitative analysis
missing in the Luria-Delbr\"uck paper and use their original data to
evaluate and compare the performance of three models: Darwinian (D),
Lamarckian (L), and Combination (C) models. The comparison is somewhat
complicated by the fact that both the L and D models are special cases
of the C model, so that C is guaranteed to fit not worse than either L
or D. We use Bayesian Model Selection
\cite{mackay1992bayesian,mackaybook}, which automatically penalizes
for more complex models (such as C) to solve the problem. We conclude
that, while the L model is certainly inconsistent with the data, D and
C explain the data about equally well when this penalty for complexity
is accounted for. Thus the Lamarckian contribution to evolution cannot
be ruled out by the 1943 Luria and Delbr\"uck data. Further, while D
and C fit equally well, neither provide a quantitatively good fit to
one of the two primary experimental data sets of Luria and
Delbr\"uck's paper, suggesting that the classic experiment may have
been influenced by factors or processes not considered.

\section{Methods}

\subsection{Models and Notational preliminaries}

There have been many theoretical attempts, with varying degrees of
success, to find closed-form analytical expressions of the
distribution of mutants under the Darwinian scenario (the Luria-Delbr\"uck distribution) following different modeling assumptions
\cite{lea1949distribution,armitage1952statistical,sarkar1991haldane,zheng2007haldane}.
For example, models with constant and synchronous division times
\cite{sarkar1991haldane,zheng2007haldane}, exponentially distributed
division times \cite{zheng1999progress,zheng2010luria}, and with
different growth rates for the wild-type and the mutant populations
\cite{koch1982mutation,jones1994luria,jaeger1995distribution} have
been proposed to find the distribution of survivors (see
Ref.~\cite{ycart2013fluctuation} for a recent review). Here we follow
Haldane's modeling hypotheses \cite{sarkar1991haldane} and assume that
(i) normal cells and mutants have the same fitness until the phage is introduced,
(ii) all cells undergo synchronous divisions,
(iii) no cell dies before the introduction of the phage,
(iv) mutations occur only during divisions, with each daughter becoming a mutant independently (D case), or only when the phage is introduced (L case), and
(v) there are no backwards mutations.

With these assumptions, the D and the C
models are able to produce very good fits to the experimental data
(see below), which suggests that relaxation of these assumptions and
design of more biologically realistic models is unnecessary in the
context of these experiments. 

For the subsequent analyses, let $N_0$ be the initial number of
wild-type, phage-sensitive bacteria, and $g$ be the number of
generations before the phage is introduced, so that the total number
of bacteria after the final round of divisions is $N=2^gN_0$, and the
total that have ever lived is $2N-N_0$. We use $\theta_D$ to denote the
probability of an adaptive (Darwinian) mutation during a division, and
$\theta_L$ to denote the probability of an adaptive Lamarckian
mutation when the phage is introduced. With this, and discounting the
probability of another mutation in the already resistant progeny, the
mean number of adaptive Darwinian mutations at generation $g$ is
$m_D=\theta_D (2N-N_0)$, and the mean number of adaptive Lamarckian
mutations is $m_L=\theta_LN$.  The number of survivors in the L model
is Poisson-distributed with the parameter $m_L$:
\begin{equation} P_L(k|\theta_L,N_0) = \frac{e^{-m_L}
    m_L^k}{k!} 
\end{equation} 

For the D model, there are multiple ways to have a certain number of
resistant bacteria, $k$, in the population of size $N$ before
introducing the phage. For instance, there are four ways to have 5
resistant bacteria (i.~e., $k=5$): (i) One mutation occurs 2 generations
before the phage introduction (where the total living population is
$N/4$ at that generation), resulting in 4 resistant progenies in the
last generation, and one more mutation at the last generation, making
a total of 5 resistant cells before the phage introduction. This is
the most likely scenario with probability
$P_{5}^{(i)}=(1-\theta_D)^{(2N-N_{0})-8} \theta_D^2 {{N/4}\choose{1}}
{{N-4}\choose{1}}$,
where $(2N-N_{0})$ is the total number of cells that have ever lived
in the entire experiment, so that $(2N-N_{0})-8$ is the total number
that have ever lived without mutating. The $\theta_D^2$ factor
indicates that a total of two mutations have occurred in the
population. The first choose factor denotes the number of independent
mutational opportunities 2 generations before the phage introduction,
and the second one denotes the number of mutational opportunities in
the last generation.  (ii) Two mutations occur 1 generation before the
phage introduction and one more mutation in the last generation. This
is less likely than (i) with probability
$P_{5}^{(ii)}=(1-\theta_D)^{(2N-N_{0})-7} \theta_D^3 {{N/2}\choose{2}}
{{N-4}\choose{1}}$,
where, there are a total of 7 mutant that have ever lived in the history of the
experiment and 3 mutational events before the introduction of the
phage.  (iii) One mutation occurs 1 generation before the phage
introduction and 3 more mutations in the last generation. This is less
likely than (ii) with probability
$P_{5}^{(iii)}=(1-\theta_D)^{(2N-N_{0})-6} \theta_D^4
{{N/2}\choose{1}} {{N-2}\choose{3}}$.
(iv) Five mutations occur in the last generation before introducing
the phage. This is the least likely scenario with probability
$P_{5}^{(iv)}=(1-\theta_D)^{(2N-N_{0})-5} \theta_D^5 {{N}\choose{5}}$.
In general, for an arbitrary number of resistant cells, $k$, let
$\Pi_{k}$ denote the set of sequences $(a_{0},a_{1},...)$ that satisfy
$k = \sum_0^{\infty} a_{s} 2^{s}$, such that
$a_s \in \Bbb Z_{\geq 0}$. This condition captures all the possible
sequences of $\big\{a_{s}\big\} \in \Pi_K$ that produce $k$ number of
resistant cells. For instance, in case (i) the corresponding sequence
is $\big\{a_{2}=1,a_{1}=0,a_{0}=1\big\}$, and in case (ii) it is
$\big\{a_{1}=2,a_{0}=1\big\}$.  Then, following Haldane's approach
\cite{sarkar1991haldane}, we can write $P_D(k)$, the probability of
finding $k$ resistant cells given the Darwinian model of evolution,
as \begin{equation} P_D(k|\theta_D,N_0) = \sum_{\big\{a_{s}\big\} \in
    \Pi_K} \left(1-\theta_D\right)^{(2N-N_{0})-\sum_{s=0}^\infty
    a_s(2^{s+1}-1)}\theta_D^x\prod_{s=0}^\infty
  {{\frac{N}{2^s} - \sum_{n=s+1}^\infty a_s(2^{n-s})}\choose{a_s}},
\label{eq:D}
\end{equation}
where $x\equiv \sum_{\big\{a_{s}\big\}}^{}a_s$ and the probability
$P_D(k|\theta_D,N_0)$ is summed over all the possible sequences
$\big\{a_{s}\big\} \in \Pi_K$ that produce the number $k$; in the case
of $P_{k=5}$ mentioned earlier,
$P_D(5|\theta_D,N_0)=P_5^{(i)}+P_5^{(ii)}+P_5^{(iii)}+P_5^{(iv)}$.

For the Combination model, both the L and the D processes contribute
to generating survivors. Thus we write the distribution of the number
of survivors in this case as a convolution \begin{equation}
  P_{C}(k|\theta_L,\theta_D,N_0) = \sum\limits_{k'=0}^{k}
  P_D(k'|\theta_D,N_0)P_L(k-k'|\theta_L,N_0). \end{equation}

Further analytical progress on the problem is hindered by additional
complications. First, in actual experiments, the initial number of
bacteria in the culture $N_0$ is random and unknown. We view it as
Poisson-distributed around the mean that one expects to have, denoted as
$\Pi(N_0|\bar{N_0})$. This gives: 
\begin{equation} P_{D/L/C}(k|\theta_D,\theta_L,\bar{N_0}) =
  \sum\limits_{N_0=0}^{\infty} P_{D/L/C}(k|N_0)\Pi(N_0|\bar{N_0}).
\end{equation}
Finally, in some of the Luria-Delbr\"uck experiments, they plated only
a fraction $r$ the entire culture subjected to the phage. This
introduced additional randomness in counting the number of survivors
after the plating, $k_p$, which we again model as a Poisson
distribution with the mean $rk$, $\Pi(k_p|rk)$
\cite{stewart1990fluctuation, stewart1991fluctuation,
  montgomery2016estimation}, resulting in the overall distribution of
survivors:
\begin{equation}
P_{D/L/C}(k_p| \theta_D,\theta_L,\bar{N_0}) = \sum_{k}^{\infty} \Pi(k_p|rk)P_{D/L/C}(k).
\label{eq:kp}
\end{equation}

\subsection{Computational models}

The expressions in the previous section appear sufficiently simple.
However, evaluating $P_C(k_p)$ in Eq.~(\ref{eq:kp}) involves two
nested sums in the expression for $P_D(k)$, Eq.~(\ref{eq:D}), a
convolution to combine L and D processes, and two more convolutions to
account for randomness in $N_0$ and during plating. These series of nested (infinite) summations make it
inefficient to use the analytical expression in Eq.~(\ref{eq:kp}) for
data analysis. Instead, we resort to numerical simulations to evaluate
$P_{D,C}(k_p)$ (expressions for the pure L model remain analytically
tractable).

Our simulations assume that each culture begins with a
Poisson-distributed number of bacteria, with a mean number of 135, as
in the original paper. The bacteria were modeled as dividing in
discrete generations for a total of $g=21$ generations. Both of the
numbers are easily inferable from the original paper using the known
growth rate and the final cell density numbers. Cells divide
synchronously, and each of the daughters can gain a resistance
mutation at division with the probability $\theta_D$, which is nonzero
in C and D models. Daughters of resistant bacteria are themselves
resistant. Non-resistant cells in the final generation are subjected
to a bacteriophage, which induces Lamarckian mutations with
probability $\theta_L$, nonzero in C and L models. We note again that
this total number of Lamarckian-mutated cells is Poisson-distributed
with the mean $\theta_L$ times the number of the remaining wild type
bacteria.

To speed up simulations of the Darwinian process, we note that the
total number of cells that have ever lived is $N_t = 2 N_02^g -
N_0$.
Thus the total number of Darwinian mutation attempts is Poisson
distributed, with mean $N_t \theta_D$. We generate the number of these
mutations with a single Poisson draw and then distribute them randomly
over the multi-generational tree of cells, marking every offspring of
a mutated cell as mutated. We then correct for overestimating the
probability of mutations due to the fact that the number of mutation
attempts in each generation decreases if there are already mutated
cells there. For this, we remove original mutations (and unmark their
progenies) at random with the probability equal to the ratio of
mutated cells in the generation when the mutation appeared to the
total number of cells in this generation. Note that since mutations
are rare, such unmarking is not very common in practice, making this
approach substantially faster than simulating mutations one generation
at a time. 

To estimate $P_C(k_p|\theta_D,\theta_L)$, we estimate this probability
on a  41x41 grid of values of $\theta_D$ and $\theta_L$. For each 
pair of values of these parameters, we perform $n=30,000,000$ simulation
runs (see below for the explanation of this choice) starting with a
Poisson-distributed number of initial bacteria, then perform
simulations as described above, and finally perform a simulated
Poisson plating of a fraction of the culture if the actual experiment
we analyze had such plating.  We measure the number of surviving bacterial cultures $k_p$ in each
simulation run and estimate $P_C(k_p|\theta_D,\theta_L)$ as a
normalized frequency of occurrence of this $k_p$ across runs,
$f_C(k_p|\theta_D,\theta_L)$. The Darwinian case is evaluated as
$P_C(k_p|\theta_D,\theta_L=0)$, and the Lamarckian case as
$P_C(k_p|\theta_D=0,\theta_L)$.

\subsection{Quality of fit}
In the original Luria and Delbr\"uck publication
\cite{luria1943mutations}, no definitive quantitative tests were done
to determine the quality of fit of either of the model to the data
well. We can use the estimated values of $P_C(k_p|\theta_D,\theta_L)$
for this task. Namely, Luria and Delbr\"uck have provided us not with
frequencies of individual values of $k_p$, but with frequencies of
occurrence of $k_p$ within bins of
$x\in(0, 1, 2, 3, 4, 5, 6-10, 11-20, 21-50, 51-100, 101-200, 201-500,
501-1000)$.
By summing $f_C(k_p|\theta_D,\theta_L) $ over $k_p\in x$, we evaluate
$f_C(x|\theta_D,\theta_L)$, which allows us to write the probability
that each experimental set of measurements, $\{n_x\}$, came from the
model:
\begin{equation}
P(\{n_x\}|\theta_D, \theta_L)={\cal C} \prod_x  \left(
  \frac{n_x}{\sum_x n_x} \right)^{f_C(x|\theta_D,\theta_L)},
\label{eq:Likelihood}
\end{equation}
where ${\cal C}$ is the usual multinomial normalization
coefficient. This probability can also be viewed as the likelihood of
each parameter combination, and the peak of the probability gives the
usual Maximum Likelihood estimation of the parameters \cite{nelsonbook}. To guarantee
that the estimated value of the likelihood has small statistical
errors, we ensured that each of the $(\theta_D,\theta_L)$ combinations
has 30,000,000 simulated cultures. Then, at parameter combinations close to the
maximum likelihood, each bin $x$ has at least 10,000
samples. Correspondingly, at these parameter values, the sampling
error in each bin is smaller than 1\%.  

Finally, to evaluate the quality of fit of a model, rather than to
find the maximum likelihood parameter values, we calculate empirically
the values of $\log_{10} P(\{n^*_x\}|\theta_D,\theta_L)$ for each
parameter combination, where $\{n^*_x\}$ are synthetic data generated
from the model with $\theta_D,\theta_L$. Mean and variance of
$\log_{10}P$ gives us the expected range of the likelihood if the
model in question fits the data {\em perfectly}.

\subsection{Comparing models}
In comparing the L, the D, and the C models, we run into the problem
that C is guaranteed to have at least as good of a fit as either D or
L since it includes both of them as special cases. Thus in order to
compare the models quantitatively, we need to penalize C for the
larger number of parameters (two mutation rates) compared to the two
simpler models. To perform this comparison, we use Bayesian model
selection \cite{mackay1992bayesian,mackaybook}, which automatically
penalizes for such model complexity.

Specifically, Bayesian model selection involves calculation of
probability of an entire model family $M=\{L,D,C\}$ rather than of its
maximum likelihood parameter values:
\begin{equation}
  P(M|\{n_x\})\propto \int
  d\vec{\theta}_MP(\vec{\theta}_{M}|\{n_x\},M) P(M)
\label{eq:BMS}
\end{equation}
where the posterior distribution of $\vec{\theta}$ is given by the
Bayes formula,
\begin{equation}
P(\vec{\theta}_{M}|\{n_x\},M)\propto P(\{n_x\}|\vec{\theta}_{M},M)P(\vec{\theta}_M|M),
\end{equation}
and $P(\{n_x\}|\vec{\theta}_M, M)$ comes from
Eq.~(\ref{eq:Likelihood}). Finally, $P(M)$ and $P(\vec{\theta}_M|M)$
are the {\em a priori} probabilities of the model and the parameter
values within the model, which we specify below.

The integral in Eq.~(\ref{eq:BMS}) is over as many dimensions as there are parameters in a given model. Thus while more complex models may
fit the data better at the maximum likelihood parameter values, a
smaller fraction of the volume of the parameter space would provide a
good fit to the data, resulting in an overall penalty on the posterior
probability of the model. Thus posterior probabilities $P(M|\{n_x\})$
can be compared on equal footing for models with different number of
parameters to say which specific model is {\em a posteriori} more
likely given the observed data. Often the integral in
Eq.~(\ref{eq:BMS}) is hard to compute, requiring analytical or
numerical approximations. However, here we already have evaluated the
likelihood of combinations of $(\theta_L,\theta_D)$ over a large grid,
so that the integral can be computed by direct summation of the
integrand at different grid points.

To finalize computation of posterior likelihoods, we must now define
the {\em a priori} distributions $P(M)$ and $P(\vec{\theta}_M|M)$. We
choose $P(C)=P(M)=P(L)=1/3$, indicating our ignorance about the actual
process underlying biological evolution. The choice of
$P(\vec{\theta}_M|M)$ is tricky, as is often the case in applications
of Bayesian statistics. We point out that the experiment was designed
so that the number of surviving mutants is almost always 1 or less, for a population with
$\approx 0.25\times10^8$ individuals, which indicated that {\em a priori} both
$\theta_L$ and $\theta_D$ are less than $4\times10^{-9}$.
Further, we assume that, for the combined model,
$P(\vec{\theta}_M)=P(\theta_L)P(\theta_D)$. Beyond this, we do not
choose one specific form of $P(\vec{\theta}_M)$, but explore multiple
possibilities to ensure that our conclusions are largely independent
of the choice of the prior.

\section{Results}

Luria and Delbr\"uck's paper provided data from multiple experiments,
where in each experiment they grew a number of cultures, subjected them to the
phage, and counted survivors. Most of the experiments have $O(10)$
cultures, which means that their statistical power for distinguishing
different models is very low. We exclude these experiments from our
analysis and focus only on experiments No.~22 and 23, which have
$n=100$ and $n=87$ cultures, respectively. The experimental protocols
differ in that Experiment 23 plated the entire culture subjected to
the phage, while Experiment 22 plated only 1/4 of the culture.  We
analyze these experiments separately from each other.

\subsection{Experiment 22}

\begin{figure}[!t]
\includegraphics[width=4in]{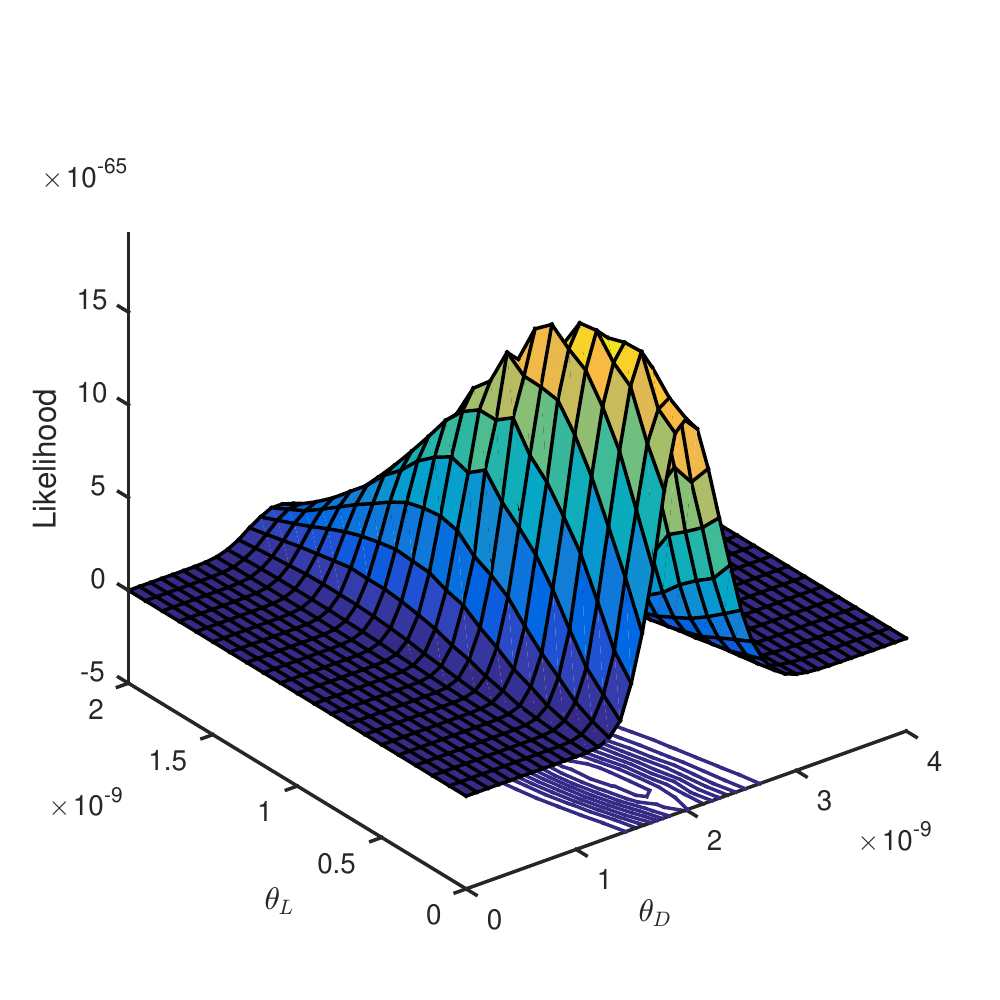}
\centering
\caption{Posterior likelihood of the Darwinian, $\theta_D$, and the
  Lamarckian, $\theta_L$, mutation parameters evaluated for the
  Luria-Delbr\"uck Experiment 22. Notice that the likelihood peaks
  away from $\theta_L=0$.\label{fig:likelihood22}} \end{figure}

\begin{figure}[!t]
\includegraphics[width=7in]{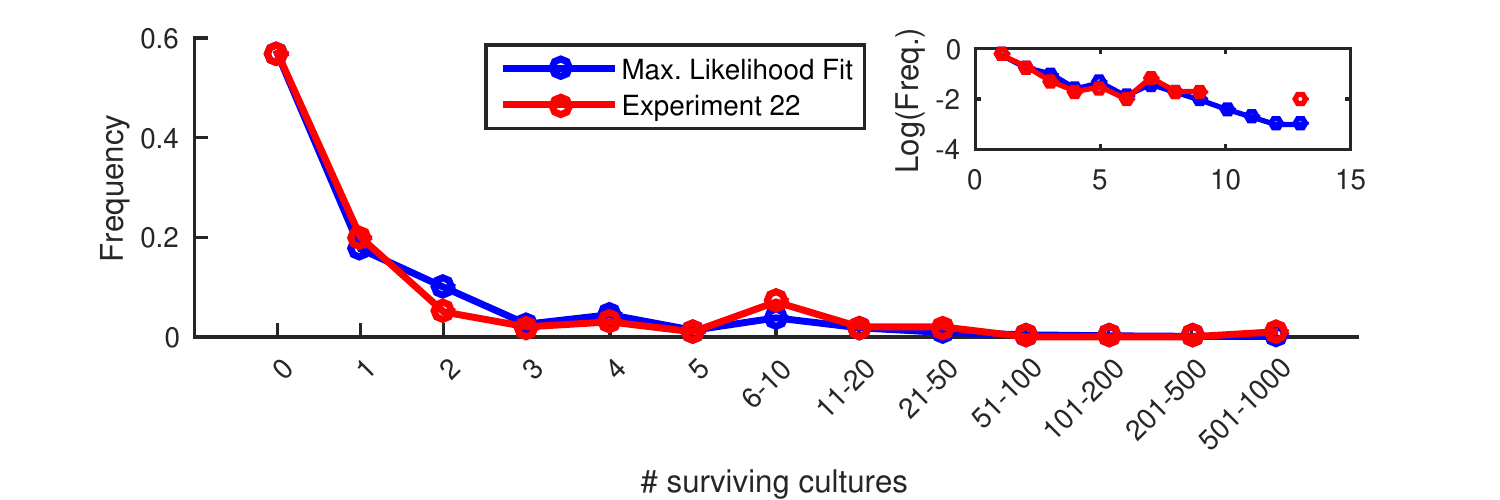}
\centering
\caption{Luria-Delbr\"uck experimental data (red) for Experiment 22
  and the maximum likelihood fit of the Combination model with
  $\theta^{22}_L=4.0 \times 10^{-10}$ and $\theta^{22}_D=1.8 \times 10^{-9}$.
  \label{fig:bins22}}
\end{figure}

We evaluated the posterior probability of different parameter
combinations, $P(\vec{\theta}_{M}|\{n_x\},M)$, numerically, as
described in {\em Methods}. The likelihood (posterior probability
without the prior term) is illustrated in
Fig.~\ref{fig:likelihood22}. Note that the peak of the likelihood is
at $\theta^{22}_L\approx4.0 \times 10^{-10} \neq 0$, $\theta^{22}_D\approx1.8 \times 10^{-9}$,  illustrating that the data suggests that the
Combination model is better than either of the pure models in
explaining the data, though the pure Darwinian model comes close.
The fit of the maximum likelihood Combination model is shown in
Fig.~\ref{fig:bins22}. The quality of the best fit
$\log_{10}P(\{n_x\}|\theta^{22}_D, \theta^{22}_L) \approx 63.7$. This
matches surprisingly well with the likelihood expected if the data was
indeed generated by the model,
$\log(P(\{n^{22}_x\}|\theta^{22}_D, \theta^{22}_L)) = 62.1\pm 5.1$. Thus the
model fits the data perfectly despite numerous simplifying
assumptions, suggesting no need to explore more complex physiological
scenarios, such as asynchronous divisions, or different growth rates
for mutated and non-mutated bacteria.

Next we evaluate the posterior probabilities of all three models by
performing the Bayesian integral, Eq.~(\ref{eq:BMS}). We use two
different priors for $\theta_L$ and $\theta_D$ to verify if our
conclusions are prior-independent: uniform between 0 and
$4\times10^{-9}$ and uniform in the logarithmic space between
$1\times10^{-10}$ and $4\times10^{-9}$. For the uniform
prior,
\begin{equation}
  \frac{P(D)}{P(C)} \approx \frac{2.8}{1}, \quad\quad
\frac{P(L)}{P(C)} \approx 10^{-10^6}.
\end{equation}
In other words, the purely Lamarckian model is ruled out by an
enormous margin, as suggested in the original publication. However,
the ratio of posterior probabilities of the Darwinian and the
Combination models is only 2.8, and this ratio is 2.0 for the
logarithmic prior, which is way over the usual 5\% significance
threshold for ruling out a hypothesis. In other words,
\begin{quote}
  The Darwinian and the Combination models of evolution have nearly
  the same posterior probabilities after controlling for different
  number of parameters in the models. Thus contribution of Lamarckian
  mechanisms to evolution in the Luria-Delbr\"uck Experiment 22
    cannot be ruled out. 
\end{quote}

\subsection{Experiment 23}

\begin{figure}[!t]
\includegraphics[width=4in]{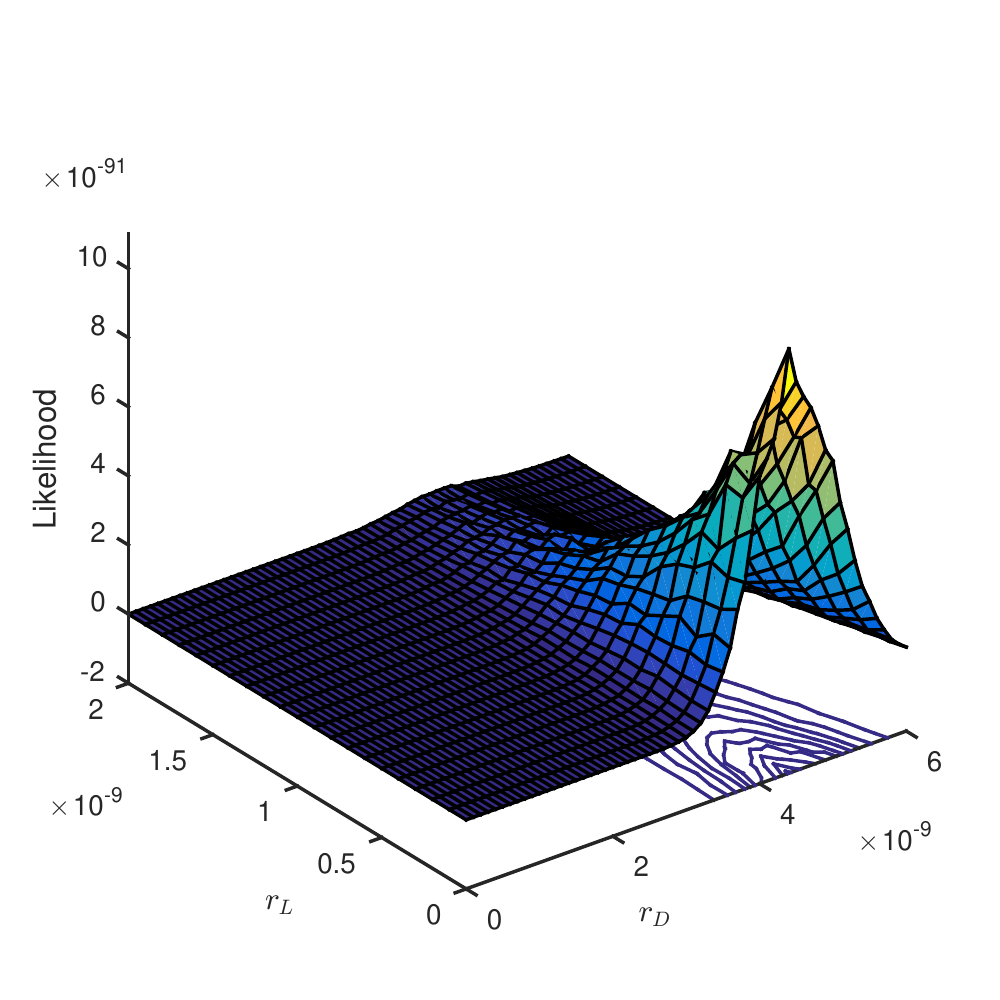}
\centering
\caption{Posterior likelihood of the mutation parameters for
  Experiment 23. The maximum likelihood is at $\theta_L=0$. However,
  neither of the three considered models is capable of fitting the
  data well (see text).}
\label{fig:likelihood23}
\end{figure}

We performed similar analysis for Experiment 23 and evaluated the
posterior likelihood, Fig.~\ref{fig:likelihood23}, for each
combination of parameters. Here, however, the posterior is several
orders of magnitude smaller than for Experiment 22. This is because
the experimental data has a tail that is {\em heavier} than typical
realizations of even the Darwinian model would predict. Indeed, Luria and
Delbr\"uck themselves noted this excessively heavy tail. However, as
they were only choosing whether the Darwinian or the Lamarckian model
fits better, this led further credence to the claim that the
Lamarckian model could not describe the data. Now we are able to
quantify this: for Experiment 23, at the maximum likelihood parameters
($\theta^{23}_L= 0,\theta^{23}_D= 4.4 \times 10^{-9}$), the quality
of fit is $\log_{10}P(\{n_x\}|\theta^{23}_D, \theta^{23}_L) \approx -90.0$.
In contrast, for data generated from the model, we get
$\log_{10}P(data|\theta_D, \theta_L) = -76.9\pm3.1$. In fact, by
generating $10^5$ data sets using these
parameter combination, we estimate that the probability of generating
data from this model that is as unlikely as the experimental data is
$p<10^{-4}$. Thus the tail of the distribution of the number of
mutants in Experiment 23 is so heavy that it cannot be fit well by
either of the hypotheses considered. Instead, it is likely that some
other dynamics are at play here, such as some form of
contamination, or additional non-Darwinian processes. In other words,
\begin{quote}
 Luria-Delbr\"uck Experiment 23 cannot be explained
by any of the proposed hypotheses (the Lamarckian, the Darwinian, or
the Combination one), and thus cannot be used to rule out one
hypothesis over another.
\end{quote}

\begin{figure}[!t]
\includegraphics[width=7in]{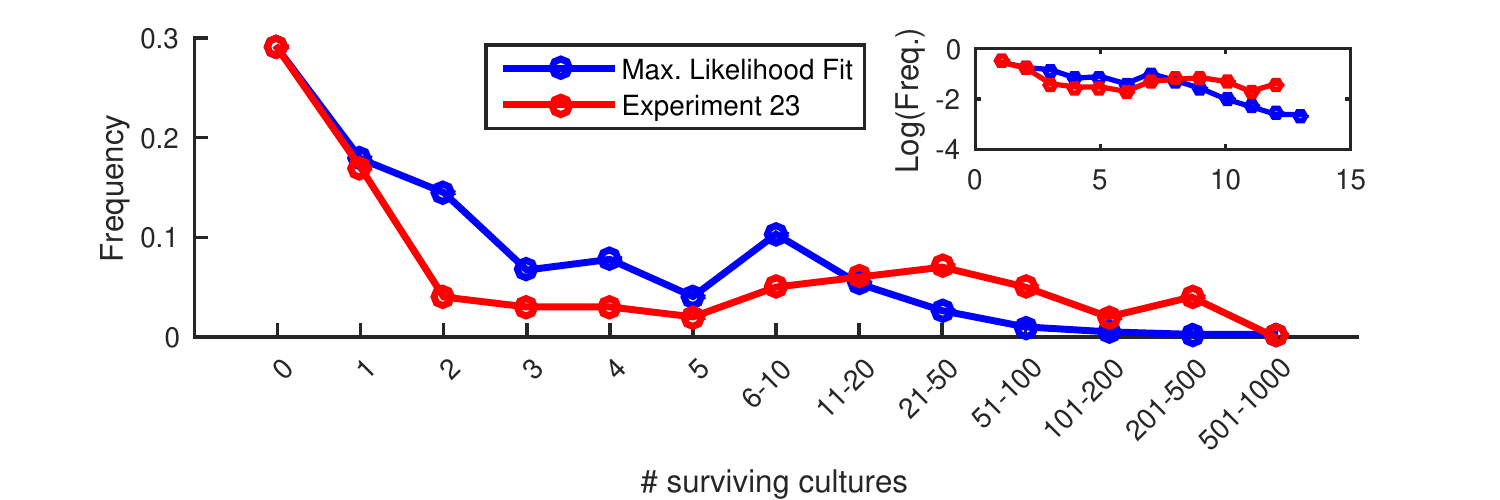}
\centering
\caption{Luria-Delbr\"uck experimental data (red) for Experiment 23
  and the maximum likelihood fit of the Darwinian model (also the best
  fit Combination model) with $\theta^{23}_D=4.4 \times 10^{-9}$. Here we can see that
  the tail in the experimental data is too heavy to be reproduced even
  by the Darwinian model. }
\label{fig:bins23}
\end{figure}

\section{Conclusion}
The classic Luria and Delbr\"uck 1943 experiment
\cite{luria1943mutations} is credited with ruling out the Lamarckian
model in favor of Darwinism for explaining acquisition of phage
resistance in bacteria. However, while heralded as a textbook example
of quantitative approaches to biology, the data in the paper was
analyzed semi-quantitatively at best.  We performed a {\em
  quantitative} analysis of the fits of three models of evolution
(Lamarckian, Darwinian, and Combination) to these classic data,
Experiments 22 and 23. Our analysis was based on a very simplified
model of the process: we started each colony with a Poisson-distributed
(mean 135) wild-type
bacteria and allowed them to replicate synchronously for exactly 21
times, with mutations occurring continuously (Darwinian model) or at
the last generation (Lamarckian model). Additionally, we did not
consider the possibility that multiple mutations might be needed to
acquire resistance, or that growth rates of bacteria may be
inhomogeneous. Nonetheless, the simple model fits Experiment 22 data
perfectly, suggesting no need for more complex modeling scenarios. 

For Experiment 22, by a ratio of $\approx 10^{-10^6}$, the Lamarckian model is
a posteriori less likely than the Darwinian one, agreeing with the
original Luria and Delbr\"uck conclusion that the pure Darwinian
evolution is a better explanation of the data than the pure Lamarckian
evolution.  However, the posterior odds of the pure Darwinian model
are only $2 - 3$ times higher than those for the Combination model
(suitably penalized for model complexity), which has nonzero Darwinian
{\em and} Lamarckian mutation rates. Even by liberal standards of
modern day hypothesis testing, there is insufficient evidence to
rule out the Combination model, and, therefore, contribution of
Lamarckian processes to bacterial evolution in this experiment. In
contrast, for Experiment 23, neither of the three considered models
could quantitatively explain the data, suggesting that additional
processes must be in play beyond simple Lamarckian and Darwinian
mutations. In summary, our analysis shows that the classic
Luria-Delbr\"uck experiments cannot be used to rule out Lamarckian
contributions to bacterial evolution in favor of Darwinism.

\section*{Acnowledgements}
This work was partially supported by NSF Grant No.~PoLS-1410978, James
S. McDonnell Foundation Grant No.~220020321, NIH NINDS R01 NS084844, Woodruff Scholars
Program and Laney Graduate School at Emory University.

\bibliographystyle{iopart-num}
%\bibliography{bibliography}   			
\providecommand{\newblock}{}

\end{document}